\UseRawInputEncoding
\documentclass[%
 reprint,
superscriptaddress,
 amsmath,amssymb,
aps,
]{revtex4-2}
\usepackage{tabularx}
\usepackage{graphicx}
\usepackage{dcolumn}
\usepackage{bm}
\usepackage{multirow}
\usepackage{xcolor}
\usepackage{hyperref}
\usepackage{soul}

\begin{document}

\preprint{APS/123-QED}
\newcommand{\bigO}{\mathcal{O}}

\title{Adaptive Neural Quantum States: A Recurrent Neural Network Perspective}

\author{Jake McNaughton}
\affiliation{Perimeter Institute for Theoretical Physics, 31 Caroline St N, Waterloo, ON N2L 2Y5, Canada}
\affiliation{Artificial Intelligence and Cyber Futures Institute, Charles Sturt University, Bathurst, NSW 2795, Australia}

\author{Mohamed Hibat-Allah}
\email{mhibatallah@uwaterloo.ca}
\affiliation{Department of Applied Mathematics, University of Waterloo, Waterloo, ON N2L 3G1, Canada}
\affiliation{Vector Institute,  Toronto,  Ontario,  M5G 0C6,  Canada}

\date{\today}
\begin{abstract}

Neural-network quantum states (NQS) are powerful neural-network ans\"atzes that have emerged as promising tools for studying quantum many-body physics through the lens of the variational principle. These architectures are known to be systematically improvable by increasing the number of parameters. Here we demonstrate an Adaptive scheme to optimize NQSs, through the example of recurrent neural networks (RNN), using a fraction of the computation cost while reducing training fluctuations and improving the quality of variational calculations targeting ground states of prototypical models in one- and two-spatial dimensions. This Adaptive technique reduces the computational cost through training small RNNs and reusing them to initialize larger RNNs. This work opens up the possibility for optimizing graphical processing unit (GPU) resources deployed in large-scale NQS simulations.
\end{abstract}

\maketitle

\section{\label{sec:level1}Introduction}

Machine learning methods are increasingly used throughout physics, ranging from experimental particle physics to quantum matter~\cite{Car19}. In particular, the intersection of machine learning and quantum simulation has emerged as a promising research direction with numerous scientific advances~\cite{dawid2025modernapplicationsmachinelearning, Melko2024}. One prominent example is Neural-network Quantum States (NQS)~\cite{ANDROSIUK1993377,LAGARIS19971,SUGAWARA2001366, carleo_solving_2017, Lan24,Medvidovic2024}, which have demonstrated state-of-the-art results~\cite{Nomura2021,hibatallah2021,chen_empowering_2024,Ren24} compared to standard numerical techniques for studying quantum many-body systems, namely Quantum Monte Carlo (QMC)~\cite{Fou01} and Density Matrix Renormalization Group (DMRG)~\cite{White92,Verstraete2023}.  

NQS is the representation of a wave function as a neural network, whose parameters are optimized through the variational principle~\cite{Mar21, McM65}, enabling a wide range of applications in quantum many-body physics~\cite{becca_sorella_2017}. In particular, finding ground states~\cite{Lan24, Medvidovic2024} and simulating time-evolution of quantum many-body systems~\cite{carleo_solving_2017, Schmitt_2020}. In the literature, a variety of neural network architectures have been used as NQSs, including Restricted Boltzmann Machines (RBM)~\cite{carleo_solving_2017,Nomura2017}, feedforward neural networks~\cite{Cai18,Choo18}, Convolutional Neural Networks (CNN)~\cite{Choo2019}, Recurrent Neural Networks (RNN)~\cite{Hibat_Allah_2020, roth_iterative_2020, Casert21, Luo23, hibatallah2024rydberg, Melko2024, Mos25, moss2025leveragingrecurrenceneuralnetwork}, and Transformers~\cite{Zhang_2023,Sprague_2024, Ren24, sobral2024physicsinformedtransformerselectronicquantum, Lange2025transformerneural}. 

Model complexity of neural networks refers to their expressive capacity, referring to their ability to approximate arbitrary functions, and is affected by a variety of factors, including the type of architecture and the number of parameters~\cite{Hu_2021_Complexity}. Different architectures contain distinctive elements, such as the hidden state in RNNs, which contribute to their complexity and whose size can be adjusted to improve the expressivity of NQS models. The latter is one key advantage of NQS architectures, compared to traditional ans\"atzes with a limited number of parameters. Specifically, their ability to be systematically improved by increasing the number of parameters, in a similar manner to the bond dimension parameter in DMRG~\cite{Schollw_ck_2011, Ganahl23}. 

With the recent considerable advances in graphical processing unit (GPU) computing, complex architectures, such as deep CNNs and Transformers, have been deployed as NQSs in quantum many-body problems~\cite{Ren24, chen_empowering_2024}. Using large models contributes significantly to the computational cost, requiring larger GPUs, more GPU units, and more time for training. As a result, despite the increased expressivity resulting from growing model complexities, most large-scale simulations addressing system-size scalability, in the literature, still rely on a small subset of NQS architectures~\cite{Nomura2021,Mos25,moss2025leveragingrecurrenceneuralnetwork, Sprague_2024}. 

To address this challenge, we demonstrate an Adaptive training framework in which the NQS model's complexity is gradually increased throughout training. More specifically, we propose an Adaptive training scheme where the dimension of the hidden state of an RNN wave function~\cite{Hibat_Allah_2020, roth_iterative_2020} is iteratively increased during training. By implementing this training scheme, higher-dimensional models are trained for a fraction of the time required to train them from scratch, thereby reducing the computational resources used.

In the machine learning community, several techniques have been developed to reduce the computational load of training neural networks, thereby enabling the training of more complex networks within existing computing resources. These schemes include transfer learning~\cite{Pan2010}, progressive neural networks~\cite{rusu2022progressiveneuralnetworks}, and Low-Rank Adaptation (LoRA)~\cite{hu2021loralowrankadaptationlarge}. Additionally, in Net2Net, methods were developed for transferring knowledge from smaller networks to larger networks, with the motivation of accelerating the exploration phase of machine learning workflows~\cite{chen2016net2netacceleratinglearningknowledge}. Furthermore, adjusting the NQS size has previously been explored in the literature through a transfer learning approach focusing on RBMs~\cite{Zen_2020, Zen21}. In Ref.~\cite{Wu_2023}, a hierarchical initialization scheme was proposed to efficiently pre-train tensorized versions of RNN wave functions~\cite{kelley_2016,Hibat_Allah_2021, hibatallah2021, Wu_2023} for a fixed hidden dimension. In contrast, our Adaptive scheme enables the scalability of the hidden dimension of RNN wave functions throughout training, improving the time efficiency and accuracy of a variational calculation across a range of testbed Hamiltonians. More specifically, our setup is comparable to the optimization scheme of Matrix Product States (MPS)~\cite{Schollw_ck_2011} with a variable bond dimension~\cite{White92, White93, Legeza_2003}, and is applicable in more than one spatial dimension. 

The plan of this paper is as follows: we introduce RNN wave functions and describe our Adaptive training scheme for improving the efficiency of the RNN variational calculations. We then share the promising results obtained from our framework applied to the 1D transverse-field Ising Model (TFIM) with nearest-neighbor interactions, the 2D Heisenberg Model, the 1D TFIM with long-range interactions, and the 1D cluster state. We demonstrate that these results indicate a superiority of the Adaptive RNN not only in terms of speed, but also in terms of achieving better accuracy and stability.

\section{Methods}

\subsection{Recurrent Neural Networks}

\begin{figure*}
    \centering
\includegraphics[width=0.9\linewidth]{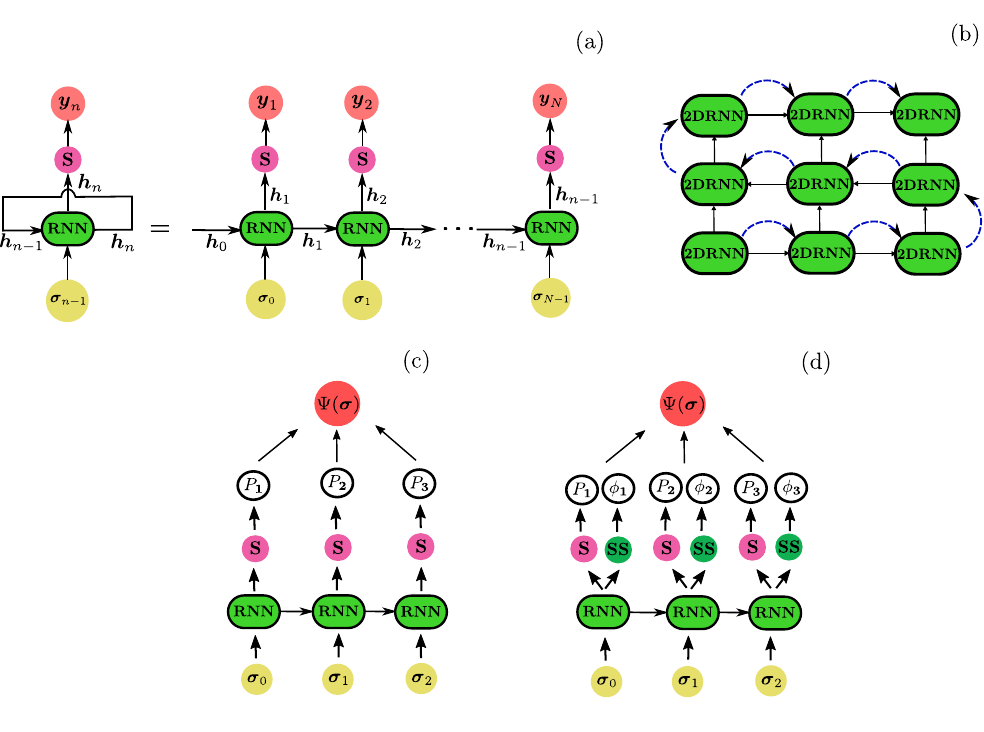}
    \caption{(a) An illustration of a recurrent neural network (RNN) in the rolled version on the left-hand side and the unrolled version on the right-hand side. Each RNN cell (in green) receives a one-hot encoding $\bm{\sigma}_{n-1}$ of the spin $\sigma_{n-1}=0,1$ in addition to a hidden state $\bm{h}_{n-1}$.   This cell outputs a hidden state $\bm{h}_n$, which is passed to a softmax layer (S in pink) which computes a two-dimensional vector $\bm{y}_n$ modeling the conditional probability of getting the next spin $\sigma_{n}$ based on the value of the previous spins. (b) A two-dimensional recurrent neural network (2D RNN) scheme, where each RNN cell received vertical and horizontal hidden states and one-hot inputs to model a two-dimensional lattice system. The zig-zag path in blue-dashed arrows provides the order of sampling. (c) An illustration of a positive RNN wave function (pRNN), where the RNN wave function is modeled as a square root of the RNN joint probability provided by the softmax layers. (d) A visualization of a complex RNN wave function (cRNN) where we model the amplitude using a square root of a probability given by the softmax layers, and a phase given by the softsign layers (SS in green).}
    \label{fig:RNN}
\end{figure*}

RNNs have enabled significant advances in natural language processing, namely in speech recognition and machine translation~\cite{lipton2015RNN}. Interestingly, these architectures are universal approximators of sequential data~\cite{Shafer2006}, simulators of Turing machines~\cite{RNNTuring}. In addition, they demonstrated strong evidence for practical use in quantum many-body physics~\cite{Carrasquilla_2019, Hibat_Allah_2020, roth_iterative_2020, hibatallah2021, Casert21, Luo23, RNN_topologicalorder, hibatallah2024rydberg, Melko2024, Mos25}. RNNs belong to the class of autoregressive models, which take advantage of the probability chain rule:
\begin{equation}
    P(\sigma_1, \sigma_2, \ldots, \sigma_N) = P(\sigma_1) P(\sigma_2 | \sigma_1) \ldots P(\sigma_N | \sigma_1, \ldots \sigma_{N-1}).
    \label{eq:chain_rule}
\end{equation}
Hereafter $(\sigma_1, \sigma_2, \ldots, \sigma_N)$ stands for a configuration of spins of size $N$ where $\sigma_n = 0, 1$. To illustrate how RNNs take advantage of the chain rule, let us take the example of the simplest RNN cell called the Vanilla RNN~\cite{lipton2015RNN}, where a spin configuration is generated sequentially through the following recursion relation:
\begin{equation}
    \bm{h}_n = f( W\bm{h}_{n-1} + V\bm{\sigma}_{n-1} + \bm{b}),
    \label{eq:recursion_relation}
\end{equation}
where $W, V$ and $\bm{b}$ are respectively the weights and the biases. $\bm{\sigma}_{n-1}$ is the one hot encoding of the spins $\sigma_{n-1}$. Furthermore, $f$ is a non-linear activation function. This computation scheme is illustrated in Fig.~\ref{fig:RNN}(a). The intialization of the recursion relation is given by $\bm{h}_0 = \bm{0}$, $\bm{\sigma}_0 = \bm{0}$. The hidden state $\bm{h}_n$ can be used to compute the parameterized conditional probability of getting $\sigma_{n}$ as:
\begin{equation}
     P_{\bm{\theta}}(\sigma_n | \sigma_{<n}) = \text{Softmax} \left ( U \bm{h}_n + \bm{c} \right ) \cdot \bm{\sigma}_n.
    \label{eq:softmax_layer}
\end{equation}
The product of the conditionals for each step $n$, allows us to obtain a parameterized joint probability distribution for the spin configurations. Note that the use of the vector $\bm{h}_n$ allows to model the conditional dependencies. For this reason, $\bm{h}_n$ is called the memory state (or the hidden state). The size of this state, called $d_{\rm h}$, controls the expressiveness of the RNN. Additionally, the RNN construction is also key for enabling perfect (autoregressive) sampling from the joint probability $P$, where $\sigma_n$ can be sampled sequentially from the conditional probabilities~\cite{Hibat_Allah_2020}. In this paper, we use a specific type of RNN cell, known as Gated Recurrent Units (GRU)~\cite{cho2014learningphraserepresentationsusing} as described in App.~\ref{app:GRUs}.

RNNs can model not only one-dimensional distributions, but can also be generalized to model two-dimensional quantum states~\cite{Hibat_Allah_2020} as illustrated in Fig.~\ref{fig:RNN}(b). Encoding two-dimensional correlations can be achieved using a two-dimensional RNN (2D RNN) through a two-dimensional recursion relation
\begin{equation*}
    \bm{h}_{i,j} = f(W[\bm{h}_{i-(-1)^j,j}; \bm{h}_{i,j-1}; \bm{\sigma}_{i-(-1)^j,j}; \bm{\sigma}_{i,j-1}]+ \bm{b}),
\end{equation*}
where $[. \; ;\; . \; ;\; . \; ;\; .]$ is a concatenation operation. Note that the previous recursion relation can be adapted to take next-nearest neighbors or other geometries into account~\cite{Hibat_Allah_2020, RNN_topologicalorder, hibatallah2024rydberg}. The 1D path for sampling and inference can be chosen as a zigzag path, as demonstrated in the dashed yellow arrows in Fig.~\ref{fig:RNN}(b). We can use the Softmax layer to compute the conditional probabilities as in the case of the 1DRNNs. For the two-dimensional benchmarks, we use a 2D GRU variant, which is explained in App.~\ref{app:GRUs}.

A quantum state amplitude $\Psi(\bm{\sigma})$ could be modeled as follows:
\begin{equation*}
    \Psi(\bm{\sigma}) = \sqrt{P(\bm{\sigma})} \exp\left({\rm i} \phi(\bm{\sigma})\right),
\end{equation*}
where $P$ is a joint probability and $\phi$ is a phase. A large family of  Hamiltonians, so-called stoquastic Hamiltonians~\cite{Bravyi:2008:CSL:2011772.2011773}, admits ground states with positive amplitudes. As a result, the ground state amplitudes can be modeled as the square root of a joint probability:
\begin{equation*}
    \Psi(\bm{\sigma}) = \sqrt{P(\bm{\sigma})},
\end{equation*}
where $P(\bm{\sigma})$ is a product of conditional probabilities computed using the RNN as illustrated in Fig.~\ref{fig:RNN}(c). This RNN wave function is denoted as a positive RNN (pRNN) wave function~\cite{Hibat_Allah_2020}. For non-stoquastic Hamiltonians, we can use a complex RNN (cRNN) wave functions~\cite{Hibat_Allah_2020}, illustrated in Fig.~\ref{fig:RNN}(d), where 
\begin{equation}
    \Psi(\bm{\sigma}) = \sqrt{P(\bm{\sigma})} \exp\left({\rm i} \phi(\bm{\sigma})\right).
\end{equation}
Here $\phi(\bm{\sigma})$ is computed as a sum of conditional phases $\phi_n$, where each $\phi_n = \phi_{\bm{\theta}}(\sigma_n | \sigma_{<n})$ is the output of a softsign layer (SS), i.e.,
\begin{equation}
     \phi_{\bm{\theta}}(\sigma_n | \sigma_{<n}) = \pi ~ \text{Softsign} \left ( U \bm{h}_n + \bm{c} \right ) \cdot \bm{\sigma}_n.
    \label{eq:softsign}
\end{equation}
Note that $\text{Softsign}(x) = x/(1+|x|)$ is chosen such that the conditional phases $\phi_n \in (-\pi, \pi)$~\cite{Hibat_Allah_2020}.

\subsection{Adaptive Recurrent Neural Networks}

\begin{figure*}
    \centering
    \includegraphics[width=0.9\linewidth]{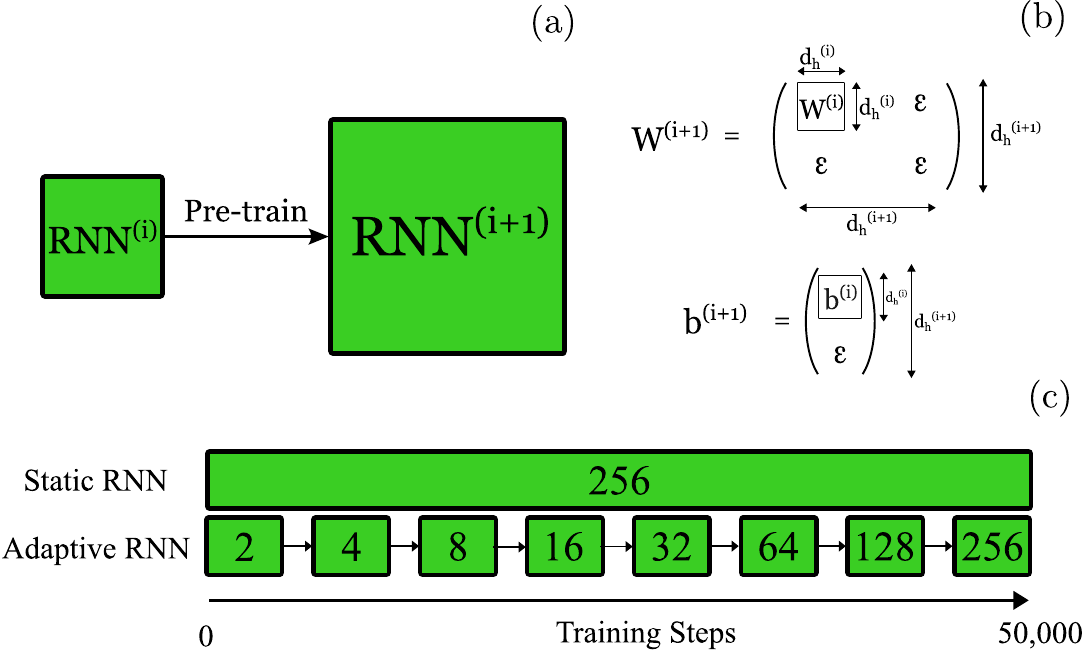}
    \caption{(a) Each RNN model (green box) in the sequence is pretrained by the previous model, and inherits the parameters. (b) When the hidden dimension increases, the weights and biases are padded with small random numbers $\epsilon$ to initialize the parameter dimensions of the next model. Note that the random numbers we use are different. (c) A diagram showing the difference between the training dynamics of a Static and Adaptive model through the example we use on the 1D TFIM. The numbers in each RNN cell indicate the RNN hidden dimension size $d_{\rm h}$.}
    \label{fig:Method}
\end{figure*}

To reduce the computational load of the model and the time taken in a variational calculation, we propose the \textit{Adaptive} RNN where the size of the hidden state is gradually increased throughout training. As a result, the dimensions of the model parameters change with the hidden state size. We develop a method to increase the size of the parameters during training and transfer them to an RNN with a larger hidden state, as illustrated in Fig.~\ref{fig:Method}(a). The goal of this Adaptive scheme is to reduce training time and improve the accuracy of variational calculations, as we demonstrate in the results section.

Our Adaptive scheme is illustrated in Fig.~\ref{fig:Method}(a). Here, when shifting from a model with hidden-state dimension $d_{\rm h}^{(i)}$ to one with dimension $d_{\rm h}^{(i+1)}$, where $d_{\rm h}^{(i)} < d_{\rm h}^{(i+1)}$, the weights and biases sizes increase from $d_{\rm h}^{(i)}\times d_{\rm h}^{(i)}$ and $d_{\rm h}^{(i)}$  to $d_{\rm h}^{(i+1)}\times d_{\rm h}^{(i+1)}$ and $d_{\rm h}^{(i+1)}$ respectively. To transfer the parameters from the smaller model to the larger model, the smaller model parameters are padded with small random numbers until they reach the appropriate dimensions for the RNN model $i+1$, as demonstrated in Fig.~\ref{fig:Method}(b).

An Adaptive RNN is a sequence of RNNs, each with a larger hidden-state dimension than the previous one. These RNNs are trained sequentially, each for a specific number of training steps. The key idea is that the final RNN model can be trained for fewer training steps as it is pre-trained by the models earlier in the sequence, which are computationally cheaper and take less time to train. In our study, we double the hidden state size after each interval $d_{\rm h}^{(i+1)} = 2 d_{\rm h}^{(i)}$. Note that we ensure that all parameters of the RNN $i+1$ are trainable, and we do not freeze the transferred set of parameters during training.

We refer to the traditional method of training RNNs with a single model of fixed hidden dimension as \textit{Static}. We term our proposed method of training, where the hidden dimension is increased throughout training, \textit{Adaptive}. This notation will be used hereafter. Fig.~\ref{fig:Method}(c) provides an example of the difference between the methods with a fixed hidden dimension of $d_{\rm h} = 256$ in the Static setup, and a hidden dimension which begins at $d_{\rm h} = 2$ and doubles at fixed intervals until reaching $d_{\rm h} = 256$ in the Adaptive setting. Note that we use this scheme when studying the one-dimensional transverse-field Ferromagnetic Ising Model (1D TFIM).

\section{Results}

To compare Static RNN and Adaptive RNN wave functions, we focus on the task of finding the ground state of several prototypical Hamiltonians. To do so, we use the Variational Monte Carlo (VMC) framework~\cite{becca_sorella_2017}, which involves minimizing the variational energy $E_\theta = \langle \Psi_{\theta} | \hat{H} | \Psi_{\theta} \rangle$ of a variational ansatz $| \Psi_{\theta} \rangle$, such as an RNN wave function, which is normalized by construction~\cite{Hibat_Allah_2020}. To find an approximation of the ground state using VMC, the parameters are learned by training the RNN parameters through a gradient descent algorithm. In this study, we use Adam optimizer~\cite{kingma2014adam} and follow the same training scheme as in Ref.~\cite{Hibat_Allah_2020}. The hyperparameters used for all benchmarks can be found in App.~\ref{app:Hyperparameters}.

\subsection{One-dimensional Transverse-field Ferromagnetic Ising Model}

To demonstrate the effectiveness of the proposed Adaptive method, a one-dimensional RNN is used to study the 1D TFIM, within open boundary conditions (OBC), described by the following Hamiltonian
\begin{equation} \label{tfim}
    \hat{H}_{\rm TFIM}=-\sum_{i=1}^{N-1} \hat{\sigma}^z_{i}\hat{\sigma}^z_{i+1}-\Gamma\sum_{i=1}^N \hat{\sigma}^x_{i}.
\end{equation}
Here $\sigma^{x,z}_{i}$ represents Pauli Matrices of the $i$th spin and $\Gamma$ is the strength of the external transverse magnetic field~\cite{Cip87}. When implementing the Adaptive method, there are various options for determining how and when to transition between models. In our 1D TFIM testbed, we experiment with the simplest approach, i.e., switching models at fixed step intervals.

A GRU-based RNN is trained on the system sizes $N = 20$, $40$, $60$, $80$, and $100$ spins at the critical point $\Gamma = 1$. Both Static and Adaptive models are trained for 50,000 gradient descent steps. The Static RNN has a hidden dimension equal to 256 throughout training, whereas the Adaptive RNN starts with $d_{\rm h} = 2$, and increases at a fixed interval (every 6,250 training iterations) as illustrated in Fig.~\ref{fig:Method}(c). A fixed learning rate of $5\times10^{-4}$ is used for all Static RNNs. This value is determined by testing a variety of learning rate experiments, which are demonstrated in App.~\ref{app:Hyperparameters}. The learning rate for the Adaptive RNNs is fixed at $5\times10^{-3}$ for the first half of training (the first 25,000 steps) and changed to $5\times10^{-4}$ for the second half.

\begin{figure}
    \centering
    \includegraphics[width=\linewidth]{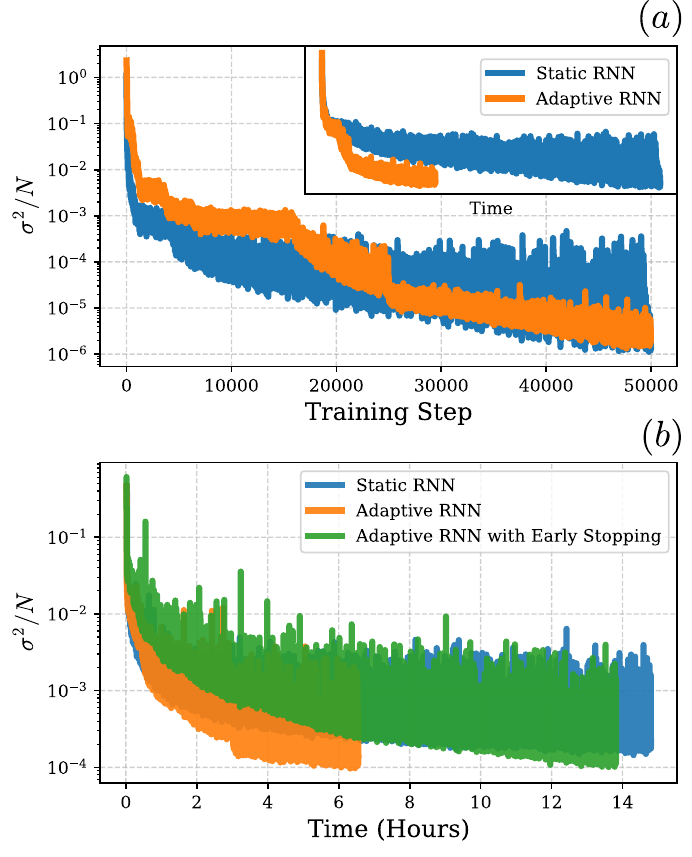}
    \caption{(a) Energy variance per spin throughout training for $N = 100$ spins in the one-dimensional transverse field Ising model. Inset: energy variance per spin against runtime. (b) Energy variance per spin vs Time (in hours) for the Heisenberg model on a lattice of 6 $\times$ 6 spins. The Static method is compared to the Adaptive method run for the same number of steps, and the Adaptive method run with early stopping.}
    \label{fig:1D_2D}
\end{figure}

Fig.~\ref{fig:1D_2D}(a) shows the variance per spin throughout training of the Adaptive and Static RNNs for $N=100$ spins. The Static RNN demonstrates a quick convergence compared to the Adaptive RNN in the first half of training. Nevertheless, in the second half for $d_{\rm h}\geq 64$, the Adaptive RNN reaches a comparable variance till the end of training. Looking at the variance evolution with time in the inset of Fig.~\ref{fig:1D_2D}(a), we observe that the Adaptive model maintains a lower variance from the beginning, and finishes training in $34\%$ of the time. This result demonstrates that the Adaptive RNN can achieve an accurate result faster compared to the Static RNN. To complement this result, we show, in App.~\ref{appendix:Time}, how the time ratio of the time taken by the Adaptive RNN over that of the Static RNN evolves with the number of spins. The ratio in the asymptotic limit is estimated around $25.6\%$ in the case of our Adaptive scheme with fixed intervals. Additionally, we report pronounced fluctuations in the Static RNN training trajectory relative to the Adaptive RNN, underscoring the improved stability achieved through the Adaptive training strategy as suggested in Fig.~\ref{fig:1D_2D}(a) and further highlighted in App.~\ref{appendix:training}.

Tab.~\ref{tab:1D_Results} presents the final results for all system sizes when computed with 1,000,000 samples after training is complete. The energy, energy variance per spin $\sigma^2/N$, relative error, and the time taken for training are provided. In addition to showing the Static and Adaptive RNN results for $d_{\rm h} = 256$, we also provide data for the trained penultimate RNN model in the Adaptive sequence with $d_{\rm h}=128$. The results clearly demonstrate that the Adaptive RNN yields compatible energies with those of the Static RNN within error bars, except for $N = 40$ where the Adaptive RNN provided the best relative error. We also note that the Adaptive RNN outperforms the Static RNN on system sizes $N = 40, 80$, and $100$ in terms of energy variance. The latter is a good indicator of the quality of a variational calculation~\cite{ becca_sorella_2017,Assaraf_2003, Wu24}. Furthermore, the penultimate model with $d_{\rm h}=128$ achieved comparable energies to the Static RNN with $d_{\rm h} = 256$, with a much shorter runtime and requiring less GPU resources. This result highlights the possibility of getting comparable accuracy with a lower number of parameters by virtue of the enhanced trainability provided by the Adaptive scheme.

\begin{table*}[]
\begin{tabular}{|c|c|c|c|c|c|}
\hline
        Method & $N$   & Energy     & $\sigma^2/N$ $[\times 10^{-6}]$           & Relative Error $[\times 10^{-7}]$  & Time (hh:mm:ss)            \\
\hline
Static RNN ($256$) & \multirow{3}{*}{20} & -25.107793(5) & \textbf{1.067(2)} & 1(2) & 00:08:44 \\
Adaptive RNN ($2\rightarrow 128$) & & -25.107794(5) &	1.203(2) & 1(2) & \textbf{00:03:59}\\
Adaptive RNN ($2\rightarrow 256$)&  & -25.107785(6) & 1.877(3) & 5(2) &  00:05:27 \\

\hline
Static RNN ($256$) & \multirow{3}{*}{40} & -50.569396(9) & 2.147(3)  & 8(1)    & 00:23:47 \\

Adaptive RNN ($2\rightarrow 128$) && -50.56941(2)&	2.144(3) & 4(1)  &  \textbf{00:07:48}\\

Adaptive RNN ($2\rightarrow 256$) &  & -50.569426(8) & \textbf{1.749(3)} & 1(2) & 00:11:12 \\

\hline
Static RNN ($256$)& \multirow{3}{*}{60} & -76.033138(9) & \textbf{1.228(2)} & 2(1) &  00:45:33 \\
Adaptive RNN ($2\rightarrow 128$) & & -76.03312(3) &	2.013(3)	& 4(1) & \textbf{00:12:14}\\
Adaptive RNN ($2\rightarrow 256$) &  & -76.03314(1) &  2.042(3) & 3(1) & 00:18:21 \\

\hline
Static RNN ($256$)& \multirow{3}{*}{80} & -101.49738(2) & 2.972(4) & 3(2) &  01:13:22 \\
Adaptive RNN ($2\rightarrow 128$) & & -101.49737(4)	& 2.190(3) & 4(1) &\textbf{00:17:25}\\
Adaptive RNN ($2\rightarrow 256$) &  & -101.49739(1) & \textbf{1.433(2)} & 2(1) &  00:27:21 \\
\hline
Static RNN ($256$)& \multirow{3}{*}{100} & -126.96182(2) & 2.313(3) & 4(1) &  01:54:56 \\
Adaptive RNN ($2\rightarrow 128$) & & -126.96184(3) & 3.638(5)	& 3(2) &  \textbf{00:24:31}\\
Adaptive RNN ($2\rightarrow 256$)&  & -126.96185(1) & \textbf{2.048(3)} & 2(1) & 00:39:46 \\

\hline

\end{tabular}

\caption{Comparison between the Static and Adaptive RNNs in terms of the final energies, variances per spin $\sigma^2/N$, and relative error for a range of system sizes using 1,000,000 samples. Times taken for training are also reported. Hereafter, the relative error is defined as $(E_{\rm RNN} - E_{\rm DMRG})/|E_{\rm DMRG}|$, where $E_{\rm DMRG}$ is energy obtained from DMRG. The best values, while taking error bars into account, for variance per spin for each system size are shown in bold. Additionally, the fastest experiments are highlighted in bold on the Time column. The error bar on the variance is estimated by assuming a Gaussian distribution over the local energies~\cite{mood1950introduction}. Note that all simulations, hereafter, were run using A100 GPUs. 
}

\label{tab:1D_Results}
\end{table*}

\subsection{Two-dimensional Heisenberg Model}
We now focus our attention on the 2D Heisenberg model on the square lattice to assess the Adaptive RNN's performance in two spatial dimensions. Historically, this model has served as a very useful playground for the development of numerical methods in computational quantum matter~\cite{sandvik_quantum_1991, liu_variational_1989, white_neel_2007, carleo_solving_2017}. The following Hamiltonian describes this model within OBC:
\begin{equation} \label{heisenberg}
    \hat{H} = \frac{1}{4}\sum_{\langle i, j \rangle} \hat{\sigma}_{i}^{x}\hat{\sigma}_{j}^{x} + \hat{\sigma}_{i}^{y}\hat{\sigma}_{j}^{y} + \hat{\sigma}_{i}^{z}\hat{\sigma}_{j}^{z}.
\end{equation}
Here $\langle i, j \rangle$ indicates that the indices being summed over are nearest neighbor pairs on the square lattice. To use a positive 2D RNN, we apply a Marshall sign rule, which is equivalent to finding the ground state of the XXZ Hamiltonian where all the off-diagonal elements are negative~\cite{marshall_antiferromagnetism_1955, capriotti_quantum_2001}.

A Static 2D RNN with $d_{\rm h}=256$ is trained for 200,000 gradient steps. In the Adaptive setting, we start with a 2D RNN that has $d_{\rm h}=32$, and double it every 50,000 steps, for a total of 200,000 steps. In addition to this setup, we also implement an Adaptive framework where $d_{\rm h}$ doubles after each early stopping criterion, given by the energy variance, is triggered until reaching a model with $d_{\rm h} = 256$ where the criterion triggers training to stop. Learning rate schedules are used for the Static and Adaptive models as highlighted in App.~\ref{app:Hyperparameters}. 

\begin{table*}[]
\begin{tabular}{|c|c|c|c|c|}
\hline
Method & Energy & $\sigma^2/N$ $[\times 10^{-4}]$ & Relative Error $[\times 10^{-5}]$ & Time (hh:mm:ss) \\
\hline
Static RNN ($256$)      & -21.7254(1)           & 2.811(4)           & 6.3(5)         & 14:48:13 \\
Adaptive RNN ($32\rightarrow 256$)   & \textbf{-21.72589(8)} & \textbf{1.775(3)}  & \textbf{4.1(4)} & \textbf{06:33:23} \\
Adaptive RNN with Early Stopping ($2\rightarrow 256$)
                & -21.72551(9)          & 2.107(3)           & 5.8(4)          & 13:49:15 \\

\hline

\end{tabular}
\caption{A comparison between the final results obtained from the Static RNN and the Adaptive RNNs with and without Early Stopping models tested on the two-dimensional square lattice Heisenberg model for a system size $N = 6 \times 6$. Each model is sampled with 1,000,000 samples after training has concluded. The runtime for each training instance is also reported. The best values are shown in bold.}
\label{tab:2D_Results}
\end{table*}

Fig.~\ref{fig:1D_2D}(b) shows the energy variance per spin throughout training for the three models trained on this benchmark task. Similar to the 1D TFIM, the Adaptive RNN completes training in less than half the time, reaching lower variances. The early stopping variant also reaches lower variances but does not provide a significant improvement in training time, as it is trained for about 540,000 steps. However, we believe that there is still room for exploring optimal stopping criteria. 

Tab.~\ref{tab:2D_Results} shows the variational energies, energy variances per spin, and relative errors for each of the three models alongside the training times. Each model is sampled to output 1,000,000 configurations once training is complete. The Adaptive model achieves the best energy, error, and variance, completing training in 6 hours and 33 minutes. The Adaptive method with early stopping outperforms the Static model in terms of energy, error, and variance, taking 13 hours and 49 minutes, whereas the Static model takes 14 hours and 48 minutes.

\subsection{Long-Range Transverse-field Ferromagnetic Ising Model}

\begin{figure}
    \centering
    \includegraphics[width=\linewidth]{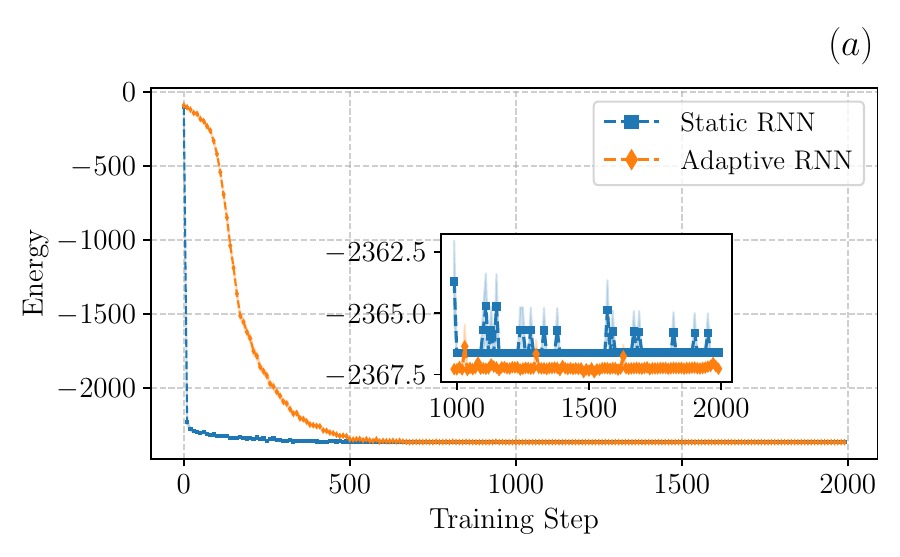}
    \includegraphics[width=\linewidth]{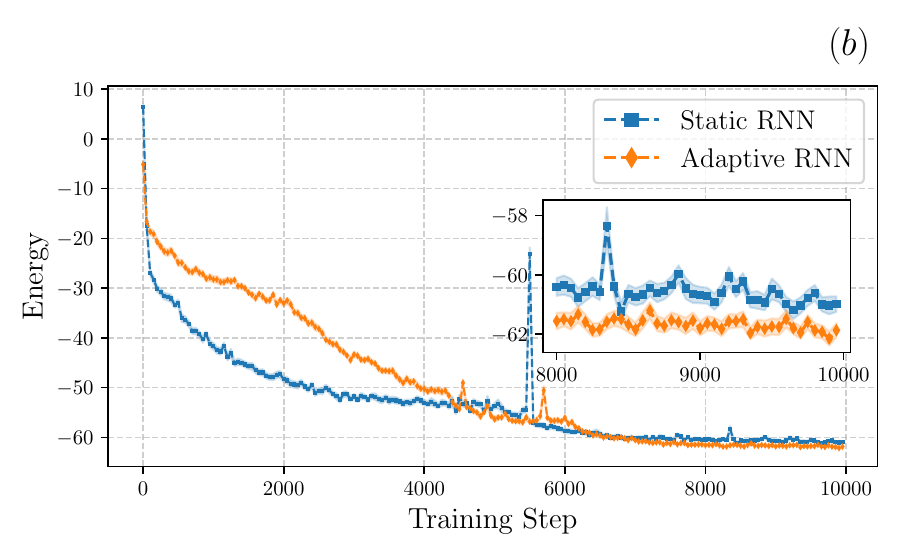}
    \caption{A comparison in terms of the variational energy between the Static RNN and the Adaptive RNN on two different models. (a) Long-range 1D TFIM  with $N = 80$ spins and $\alpha = 0.1$. Here, the Adaptive RNN size is changed every $200$ steps. (b) 1D Cluster state for $N = 64$ spins. Note that the Adaptive RNN size is changed every $1000$ steps. For both panels, the lower the energy, the better. We observe a lower variational energy for the Adaptive RNN at convergence for the two Hamiltonians.}
    
    \label{fig:Ising_and_Cluster}
\end{figure}

In the two previous benchmarks, we confirm that the Adaptive RNN can provide accurate ground state energy approximations within a shorter time frame than the Static RNN for the 1D TFIM and the 2D Heisenberg models. We now focus on the influence of our Adaptive scheme when training on long-range models. We start with the long-range TFIM given by the following Hamiltonian:
\begin{equation} \label{ILR}
    \hat{H}_{\rm LR-TFIM}=-\sum_{1\leq i<j\leq N}\frac{1}{|i-j|^\alpha}  \hat{\sigma}^z_{i}\hat{\sigma}^z_{j}-\Gamma\sum_{i} \hat{\sigma}^x_{i},
\end{equation}
where $\alpha$ is a tunable parameter. This model has been experimentally realized using trapped ion quantum simulators, where the interaction strength decays with distance as a power law $ J_{ij} \propto 1/|i - j|^\alpha $, with tunable $\alpha$~\cite{Zhang2017}. Additional realizations have been achieved using Rydberg atom arrays, allowing exploration of constrained and long-range Ising-type interactions~\cite{Bernien2017}. Additionally, this model corresponds to the short-range 1D TFIM, introduced earlier, when $\alpha \to \infty$. In our simulations, we choose $\alpha=0.1, \Gamma = 1$, and $N = 80$ spins as a playground for comparing our Adaptive RNN against the Static RNN.

Our results are summarized in Fig.~\ref{fig:Ising_and_Cluster}(a). Here we observe that near convergence, the Adaptive RNN energy $-2367.244(2)$ is lower than the Static RNN energy $-2366.55(2)$ with a difference of about $0.69(2)$, even though the Static RNN converged faster in the initial phase of training. This result suggests that Adaptive RNNs are more effective at circumventing excited states compared to Static RNNs. One possible explanation for this result can be related to Adaptive RNNs starting with a small hidden dimension, reflecting a low entanglement structure. This property is an agreement with the ground state of Ising long range, for $\alpha = 0.1$, belonging to the mean-field universality class~\cite{Koffel_2012,Defenu_2023}. Static RNNs, on the other hand, start with a large hidden dimension which are likely biased towards learning entangled states, as suggested by the RNN entanglement area law~\cite{Levine19,yang2024classicalneuralnetworksrepresent}. This finding, along with the previous benchmarks, highlights the importance of starting with a small hidden dimension at the beginning of a variational calculation to learn simple quantum states in the initial phase of training.

We also note that the runtimes for both RNNs are comparable ($\sim 12$ minutes for both Static and Adaptive RNNs). Note that the constant compilation overhead in the Adaptive RNN when switching from one model to the next is the main factor behind the comparable runtimes. However, we highlight that the Adaptive RNN can reach better energies compared to the Static RNN around halfway through training (see inset of Fig.~\ref{fig:Ising_and_Cluster}(a)), which corresponds to just $5$ minutes and a half of runtime.

\subsection{1D Cluster State}

We now shift our attention to ground states with a sign structure. In particular, we confirm a similar observation to the previous benchmark using a cRNN instead of a pRNN. To achieve this goal, we study the 1D Cluster State Hamiltonian:
\begin{align}
    \hat{H}_{\mathrm{Cluster}} &=-\sum_{k=2}^{n-2} X_{k-1} Z_k X_{k+1} \nonumber  \\ &- Z_1 X_2 - X_{n-1} X_n - X_{n-2} Z_{n-1} Z_n.
\end{align}
This Hamiltonian is a prototypical model for measurement-based quantum computation, where entanglement is generated via multi-qubit stabilizer terms (e.g. $X_{k-1} Z_k X_{k+1}$) rather than dynamic evolution \cite{raussendorf2002computationalmodelunderlyingoneway}. Its ground state belongs to a gapped, symmetry-protected topological (SPT) phase representative in 1D, with nontrivial edge modes and robustness under certain perturbations~\cite{Doherty_2009}. Note that this Hamiltonian is non-stoquastic. As a result, we use a cRNN to model the phase of its ground state~\cite{Hibat_Allah_2020}.

We focus our comparison between the Static RNN and the Adaptive RNN on this Hamiltonian with $N = 64$ spins, adopted in Ref.~\cite{yang2024classicalneuralnetworksrepresent}, for a $y$-rotation angle $\theta = 0$, such that $R_y^\dagger(\theta) \hat{H}_{\mathrm{Cluster}} R_y(\theta) = \hat{H}_{\mathrm{Cluster}}$ where $R_y(\theta)$ is the $y$-rotation unitary. This point has a ground state with the largest conditional mutual information (CMI)~\cite{yang2024classicalneuralnetworksrepresent}, indicating long-range conditional correlations. As a result, $\theta=0$ is the hardest point to learn by the RNN~\cite{yang2024classicalneuralnetworksrepresent}. Similar to the long-range TFIM model, the runtime for both Adaptive and Static RNNs is similar (around $18$ minutes for both), however we note that the Adaptive RNN takes only $13$ minutes (around $8000$ training steps) to outperform the Static RNN variational energies as illustrated in the inset of Fig.~\ref{fig:Ising_and_Cluster}(b).

Our results, illustrated in Fig.~\ref{fig:Ising_and_Cluster}(b), demonstrate a noticeable difference at the first decimal point between the Static RNN and the Adaptive RNN despite the faster convergence of the Static RNN. This result confirms once again the ability of the Adaptive RNN to better avoid local minima in the VMC optimization landscape. Additionally, even though we do not obtain the true ground state energy $-64$~\cite{yang2024classicalneuralnetworksrepresent}, our Adaptive RNN energy is within a relative error of $3.3 \times 10^{-2}$, which is smaller than that of our Static RNN ($5.0 \times 10^{-2}$) and less than half of the relative error obtained by the Static RNN in Ref.~\cite{yang2024classicalneuralnetworksrepresent}. These results also highlight the advantage provided by the Adaptive scheme in the presence of a non-trivial sign structure in the ground state. The latter is known to induce a rugged optimization landscape~\cite{Bukov_2021}, and our results suggest that the Adaptive training scheme is better equipped to navigate such landscapes.

\section{Conclusion}
In this paper, we propose a framework for training RNN wave functions by gradually scaling up the hidden state dimension throughout training. This technique resulted in a significant reduction in the time taken for training when applied to prototypical spin models studied, while reaching similar or improved levels of accuracy. Our study also demonstrates that our Adaptive RNN can reach accurate energies using a lower hidden state dimension, highlighting the improved trainability using our Adaptive scheme. Additionally, using lower-dimensional models earlier in training allows for capturing low-entangled states, such as in the case of the ground state of the long-range TFIM model.

Our study focuses on a regular schedule for growing the RNN size. However, an optimal early stopping mechanism is expected to improve the performance of the Adaptive framework by ensuring each model in the sequence is trained for long enough to gain the time advantage, and not overtrained when a greater benefit would be gained by switching to the next model. Additionally, developing an adaptive learning rate scheme that depends on the stage of our Adaptive method can improve training and speed of convergence. Furthermore, combining our Adaptive RNN with the iterative retraining technique of RNNs~\cite{roth_iterative_2020,hibatallah2021,Mos25,moss2025leveragingrecurrenceneuralnetwork} will also allow targeting large lattice sizes using a fraction of the computational cost, leveraging the inherent weight sharing in RNNs. The latter provides a key advantage of Adaptive RNNs compared to Adaptive RBMs used in Refs.~\cite{Zen_2020,Zen21}. We also highlight that variational energies obtained in this work could be further improved by applying tensorization~\cite{hibatallah2021,Wu_2023, Mos25} and leveraging symmetries~\cite{Hibat_Allah_2020, hibatallah2021, Nomura_2021}.

To conclude, the proposed method of increasing the complexity of the model throughout training can be expanded in multiple directions. While we have restricted ourselves to applying Adaptive RNNs to many-body quantum systems, the Adaptive training framework could be applied to a wide variety of NQSs to reduce runtime and improve the accuracy of quantum many-body simulations with NQSs. More broadly, this framework can be adopted to improve the trainability of machine learning architectures in a wide range of applications beyond quantum many-body physics.

\begin{acknowledgments}
Computer simulations were made possible thanks to the Digital Research Alliance of Canada and the Math Faculty Computing Facility at the University of Waterloo. M.H acknowledges support from Natural Sciences and Engineering Research Council of Canada (NSERC), and the Digital Research Alliance of Canada. Research at Perimeter Institute is supported in part by the Government of Canada through the Department of Innovation, Science and Economic Development and by the Province of Ontario through the Ministry of Colleges and Universities.
\end{acknowledgments}

\section*{Code Availability}
Our implementation of the presented methods and all
scripts needed to reproduce our results in this
manuscript are openly available on GitHub \url{https://github.com/jakemcnaughton/AdaptiveRNNWaveFunctions/}.

\appendix

\section{Gated Recurrent Units (GRU)} \label{app:GRUs}
In this paper, we use Gated Recurrent Units (GRU) to implement our one- and two-dimensional RNNs~\cite{Hibat_Allah_2020}. In the one-dimensional case, we use the standard implementation of GRUs provided in Ref.~\cite{cho2014learningphraserepresentationsusing}. In the one-dimensional case, at each step $n$, the hidden state $\bm{h}_n$ is computed via a gating mechanism that interpolates between the previous hidden state $\bm{h}_{n-1}$ and a candidate state $\bm{\tilde{h}}_n$. This interpolation is governed by an update gate $\bm{u}_n$, which controls how much of the new candidate information is integrated. This gating mechanism helps mitigate the vanishing gradient problem in recurrent architectures~\cite{zhou2016minimal,shen2019mutual}. The GRU update equations are as follows:
\begin{align*}
\label{eq:GRU}
\bm{u}_n &= \text{sigmoid}\left( W_{\rm g} [\bm{h}_{n-1} ; \bm{\sigma}_{n-1} ] + \bm{b}_{\rm g} \right), \\
\nonumber
\bm{r}_n &= \text{sigmoid}\left( W_{\rm r} [\bm{h}_{n-1} ; \bm{\sigma}_{n-1} ] + \bm{b}_{\rm r} \right), \\
\nonumber
\bm{\tilde{h}}_n &= \tanh\left( \bm{r}_n \odot (W_{\rm h} \bm{h}_{n-1} + \bm{b}_{\rm h}) + W_{\rm in} \bm{\sigma}_{n-1} + \bm{b}_{\rm in} \right), \\
\nonumber
\bm{h}_n &= (1 - \bm{u}_n) \odot \bm{h}_{n-1} + \bm{u}_n \odot \bm{\tilde{h}}_n.
\end{align*}
Here, `$\odot$' denotes the element-wise (Hadamard) product, and `sigmoid' and `tanh' refer to the standard activation functions. The reset gate $\bm{r}_n$ controls how much of the past information (i.e., $\bm{h}_{n-1}$) is used when computing the candidate hidden state. Note that the weight matrices $W_{\rm g},  W_{\rm r}, W_{\rm h}, W_{\rm in}$ and biases $\bm{b}_{\rm g}, \bm{b}_{\rm r}, \bm{b}_{\rm h}, \bm{b}_{\rm in}$ are trainable parameters of the one-dimensional GRU cell.

In the two dimensional case (2D RNN)~\cite{Hibat_Allah_2020}, to compute the hidden state $\bm{h}_{i,j}$, we first construct a candidate hidden state $\tilde{\bm{h}}_{i,j}$ based on a summary of neighboring hidden states and inputs. An update gate $\bm{u}_{i,j}$ then determines how much of this candidate state is incorporated into the final hidden state versus how much of the neighboring hidden state information is retained. The two-dimensional recursion relation is defined as follows~\cite{RNN_topologicalorder,RNN_RydbergKagome}:
\begin{align*}
    \tilde{\bm{h}}_{i,j} &= \tanh\! \Big(
    W [\bm{h}'_{i,j} ; \bm{\sigma}'_{i,j} ]
    +  \bm{b} \Big), \\
   \bm{u}_{i,j} &= \text{sigmoid}~\! \Big(
    W_{\rm g} [ \bm{h}'_{i,j} ; \bm{\sigma}'_{i,j} ]
    +  \bm{b}_{\rm g} \Big), \\
    \bm{h}_{i,j} &= \bm{u}_{i,j} \odot \bm{\tilde{h}}_{i,j} + (1-\bm{u}_{i,j}) \odot (U \bm{h}'_{i,j}).
\end{align*}
Here `$\odot$' denotes the element-wise (Hadamard) product. The vector $\bm{h}'_{i,j}$ is a concatenation of the neighbouring hidden states $\bm{h}_{i-(-1)^j,j}, \bm{h}_{i,j-1}$. The same definition also holds for $\bm{\sigma}'_{i,j}$. Note that the index $i-(-1)^j$ is used to ensure compatibility of the two-dimensional recursion relation with the zigzag sampling path. The weight matrices $W, W_{\rm g}, U$ and biases $\bm{b}, \bm{b}_{\rm g}$ are trainable parameters of the two-dimensional GRU cell.  We finally note that, before applying the Softmax layer, we apply a gated linear unit (GLU) layer~\cite{dauphin2017languagemodelinggatedconvolutional,shazeer2020gluvariantsimprovetransformer} on the hidden state as follows: 
\[\mathbf{h}_{i,j}' = (W_1 \mathbf{h}_{i,j}+\mathbf{b}_1)\odot \text{sigmoid}(W_2 \mathbf{h}_{i,j} + \mathbf{b}_2),\]
where the weights $W_1, W_2 \in \mathbb{R}^{d_{\rm h} \times d_{\rm model}}$ and biases $\mathbf{b}_1,\mathbf{b}_2 \in \in \mathbb{R}^{d_{\rm model}}$. Note that $d_{\rm model}$ is a hyperparameter that we choose as $d_{\rm model} = d_{\rm h}$.

\section{Hyperparameters} \label{app:Hyperparameters}
Tab.~\ref{tab:hyperparams} summarizes the hyperparameters used for training the different models on all benchmark Hamiltonians. Note that we trained the Static and Adaptive RNNs for the same number of steps, except for the early stopping variant of the Adaptive scheme. The early stopping variant is trained until the criterion triggers a stop, i.e., it runs for a variable number of epochs.

Adam optimizer~\cite{kingma2014adam} is used as the standard parameter optimizer in all benchmarks. This choice requires maintaining momentum throughout the Adaptive training setup by carrying an optimizer state. To maintain the information from the smaller model faithfully, the information in the optimizer state is carried over to the new model. This step requires encapsulating this data into a higher-dimensional optimizer state. When progressing from one model in the sequence to the next, we carry over both the parameters and momentum states.

\def\arraystretch{1.25}%
\begin{table*}[p]
    \centering
    \footnotesize
    \begin{tabular}{|c|c|c|c|}\hline
        Benchmark & Model & Hyperparameter & Value \\\hline
        \multirow{13}{*}{1D TFIM}    & \multirow{6}{*}{Static }& Architecture & 1D pRNN with fixed $d_{\rm h}$\\
                                    && Number of samples &  100\\
                                    && Training iterations & $50,000$  \\
                                    && Learning rate & $5\times 10^{-4}$\\
                                    && System Sizes &  20, 40, 60, 80, 100  \\
                                    && $d_{\rm h}$       & 256 \\
                        
                                    \cline{2-4}
                                    & \multirow{7}{*}{Adaptive}& Architecture & 1D pRNN with $d_{\rm model}$ and $d_{\rm h}$ doubling every $6250$ steps\\
                                    && Number of samples & 100\\
                                    && Training iterations & $50,000$  \\
                                    && Learning rate & $5\times 10^{-3}$ until 25,000 steps, then $5\times 10^{-4}$\\
                                    &&  System Sizes &  20, 40, 60, 80, 100   \\
                                    && Starting $d_{\rm h}$ & 2 \\
                                    && Final $d_{\rm h}$ & 256\\
                                   
        \hline
        \multirow{23}{*}{2D Heisenberg}  & \multirow{6}{*}{Static }& Architecture & 2D pRNN with fixed $d_{\rm model}$ and $d_{\rm h}$\\
                                        && Number of samples & 500                  \\
                                        &&  Training iterations & $200,000$         \\
                                        && Learning rate & $5\times 10^{-4}\times\left(1+\frac{t}{5000}\right)^{-1}$                         \\
                                        && System Sizes & $6\times 6$               \\
                                        &&   $d_{\rm h}$  &  256           \\
                                        \cline{2-4}

                                        & \multirow{8}{*}{Adaptive }& Architecture & 2D pRNN with $d_{\rm h}$ doubling every 25,000 steps       \\
                                        && Number of samples &      500                  \\
                                        &&  Training iterations &   200,000           \\
                                        && Learning rate &          $5\times 10^{-4}\times\left(1+\frac{t-100,000}{5000}\times\left\lfloor{\frac{t}{100,000}}\right\rfloor\right)^{-1}$\\
                                        && System Sizes   &         $6\times 6$            \\
                                        && Starting $d_{\rm h}$ & 32 \\
                                        && Final $d_{\rm h}$ & 256\\
                                    
                                        \cline{2-4}
                                        & \multirow{9}{*}{Early Stopping }& Architecture & 2D pRNN with $d_{\rm h}$ doubling from early stopping \\
                                        && Number of samples &          500                         \\
                                        && Training iterations &        Variable                    \\
                                        && Learning rate &              $0.01$ until $d_{\rm h}=64$ then $0.0001$                            \\
                                        && Starting $d_{\rm h}$ & 2 \\
                                        && Final $d_{\rm h}$ & 256\\
                                        && Early Stopping Criterion &   Variance                    \\
                                        && $\delta$                    &   $10^{-\frac{1}{2}\log_{2}(d_{\rm h})}$            \\
                                        && Patience                 &   10000                            \\
                            
        \hline
             
        \multirow{10}{*}{Long-range TFIM}    &\multirow{5}{*}{Static}  &  Architecture & 1D pRNN with fixed $d_{\rm h}$       \\
                            && Number of samples & 500                  \\
                            &&  Training iterations & 2,000           \\
                            && Learning rate &  $10^{-3}$\\
                            &&   $d_{\rm h}$  &  256             \\
                            \cline{2-4}
                           & \multirow{6}{*}{Adaptive }& Architecture & 1D pRNN with $d_{\rm h}$ doubling every 200 steps       \\
                            && Number of samples & 500                  \\
                            &&  Training iterations & 2,000           \\
                            && Learning rate &  $10^{-3}$\\
                            && Starting $d_{\rm h}$ & 2 \\
                                && Final $d_{\rm h}$ & 256\\
        \hline
        \multirow{10}{*}{Cluster State}    &\multirow{5}{*}{Static}&   Architecture & 1D cRNN with fixed $d_{\rm h}$      \\
                            && Number of samples & 100                  \\
                            &&  Training iterations & 10,000           \\
                            && Learning rate &  $10^{-4}$\\
                            &&   $d_{\rm h}$  &  256             \\
    \cline{2-4}
    & \multirow{6}{*}{Adaptive}      & Architecture & 1D cRNN with $d_{\rm h}$ doubling every 1,000 steps       \\
                            && Number of samples & 100                  \\
                            &&  Training iterations & 10,000           \\
                            && Learning rate &  $10^{-3}$\\
                             && Starting $d_{\rm h}$ & 2 \\
                                && Final $d_{\rm h}$ & 256\\
        \hline
    \end{tabular}
    \caption{A summary of the Hyperparameters used on the four different benchmarks in this Paper.}
    \label{tab:hyperparams}
\end{table*}

\begin{figure*}
    \centering
    \includegraphics[width=\linewidth]{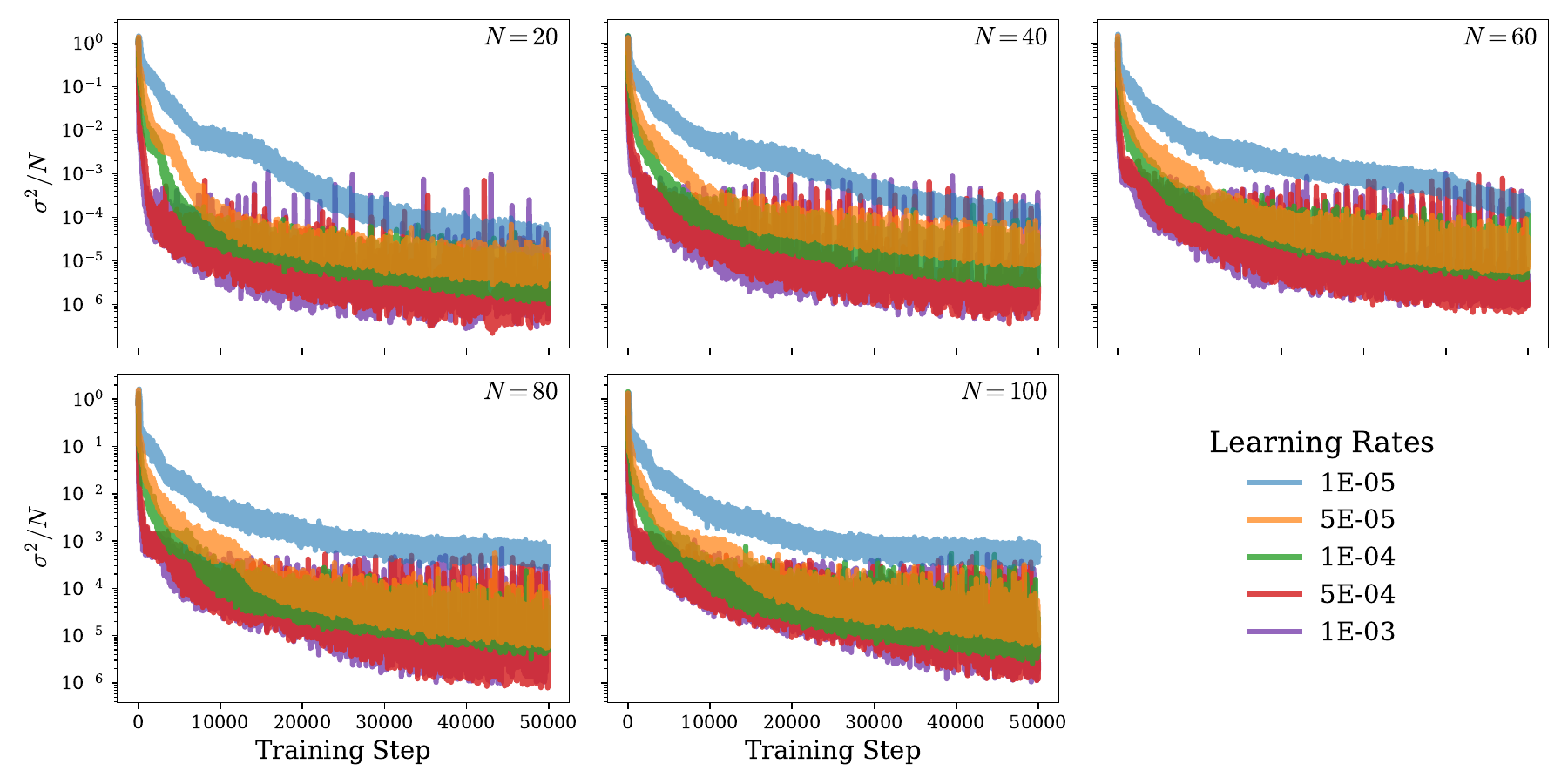}
    \caption{Variance throughout training of the Static framework for all system sizes of the 1D TFIM that were studied. Five learning rates were trialed, and $10^{-4}$ is identified as the optimal rate.}
    \label{fig:TradLR}
\end{figure*}

To determine the optimal learning rate for the Static RNNs, a range of learning rates is tested. Fig.~\ref{fig:TradLR} shows the variance throughout training of five different learning rates between $10^{-3}$ and $10^{-5}$ for each of the system sizes we studied of the 1D TFIM. For each system size, $5\times10^{-4}$ is the learning rate that achieves the lowest variances and is therefore chosen as the optimal learning rate. For the Adaptive model, the small RNNs can be trained with a larger learning rate compared to the higher-dimensional RNNs. Therefore, we change the learning rate halfway through training. This step balances the goal of simplicity (not introducing many hyperparameters) with making the most of the smaller RNNs. A variety of pairs of learning rates (for the first and second half) were trialed, and the optimal configuration is found to be a learning rate of $0.005$ for the first half of training, then changing to $0.0005$ for the remaining phase of training.

For the 2D Heisenberg model, the RNNs (both Static and Adaptive) are harder to optimize compared to the previous one-dimensional benchmark. Therefore, a learning rate schedule is used. For the Static model, an exponential decay schedule is used (see Tab.~\ref{tab:hyperparams}). To get the most out of the small models in the Adaptive framework, the first half of training is conducted at a fixed learning rate of  $5\times10^{-4}$, and a decay schedule is used for the second half of training.  For the Early Stopping setup, the learning rate is changed halfway through, similarly to the Adaptive model in the 1D TFIM experiments. Note that Early Stopping is a technique to stop a machine learning model's training once a monitored metric has ceased to improve, given a specific criterion. Three hyperparameters are required for the early stopping algorithm: the metric being monitored, the minimum change required to be considered as improving ($\delta$), and the number of consecutive epochs required for the metric to not improve in order for the early stopping to be triggered (patience). In our study, the metric is chosen as the moving average of the energy variances. The moving average, over a window of $500$ training steps, is used to reduce statistical fluctuations on the metric.

Concerning the long-range TFIM benchmark, the Static and Adaptive RNNs were trained using three different learning rates: $10^{-4}$, $5 \times 10^{-4}$, and $10^{-3}$, and we report the best results corresponding to $10^{-3}$ for both RNNs. Similarly, for the 1D Cluster Hamiltonian, we train the Static and Adaptive RNN on the same set of learning rates, and we find that the best result corresponds to a learning rate of $10^{-4}$ for the Static RNN and $10^{-3}$ for the Adaptive RNN.

\section{Analysis of Time} \label{appendix:Time}
As the size of the system increases, the ratio of the time taken for training between the Adaptive and Static RNNs decreases. The time taken to train an RNN wave function increases quadratically as the system size increases. Here, we provide a scaling study of the ratio to demonstrate the improvement that the Adaptive framework can achieve when applied to large system sizes. We perform this analysis on the data we collected for the 1D TFIM, where we study the largest number of different system sizes.

\begin{figure}
    \centering
    \includegraphics[width=\linewidth]{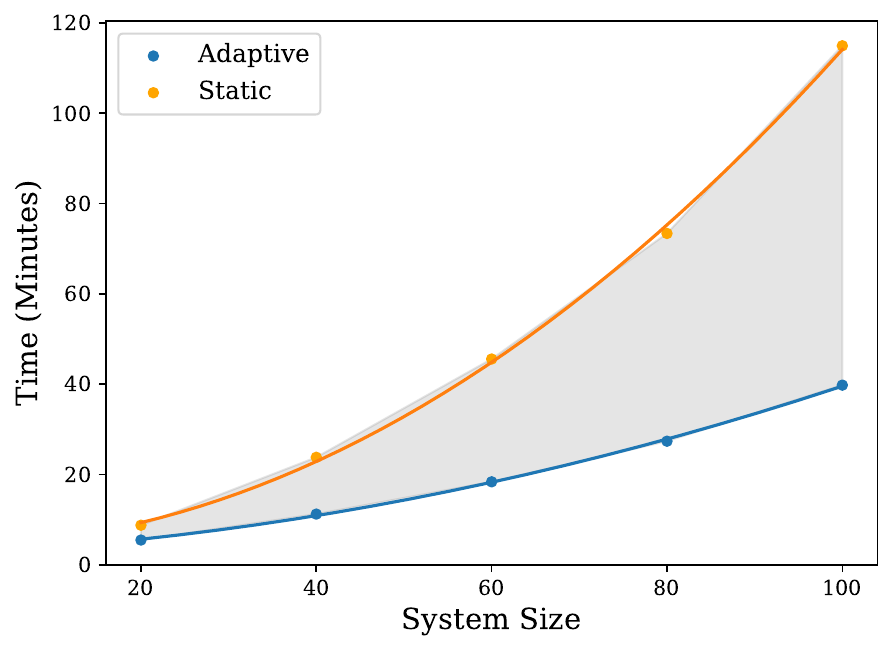}
    \caption{Comparison of time taken for training the Atatic and Adaptive RNN using the 1D TFIM as a testbed. The parabolic curves are fitted to both sets of data. Grey background shows the increasing difference between times as the system size increases.}
    \label{fig:TFIM_runtime}
\end{figure}

Fig.~\ref{fig:TFIM_runtime} shows the time taken for both models, with parabolas fitted using SciPy Optimize. The fits to the Adaptive runtime is given by $$T_{\rm Adaptive}(N)=0.00271N^2+0.0986N+2.58$$ and the fit for the Static runtime is $$T_{\rm Static}(N)=0.0106N^2+0.0438N+4.22.$$ Therefore the limit of the ratio is given by $$\lim_{N\to\infty} \frac{T_{\rm Adaptive}}{T_{\rm Static}}=\frac{0.00271}{0.0106}=25.6\%,$$ indicating that as the system size increases, the Adaptive model takes approximately a quarter of the time the Static model takes to train.

\section{Stability of Training} \label{appendix:training}

\begin{figure*}
    \centering
    \includegraphics[width=\linewidth]{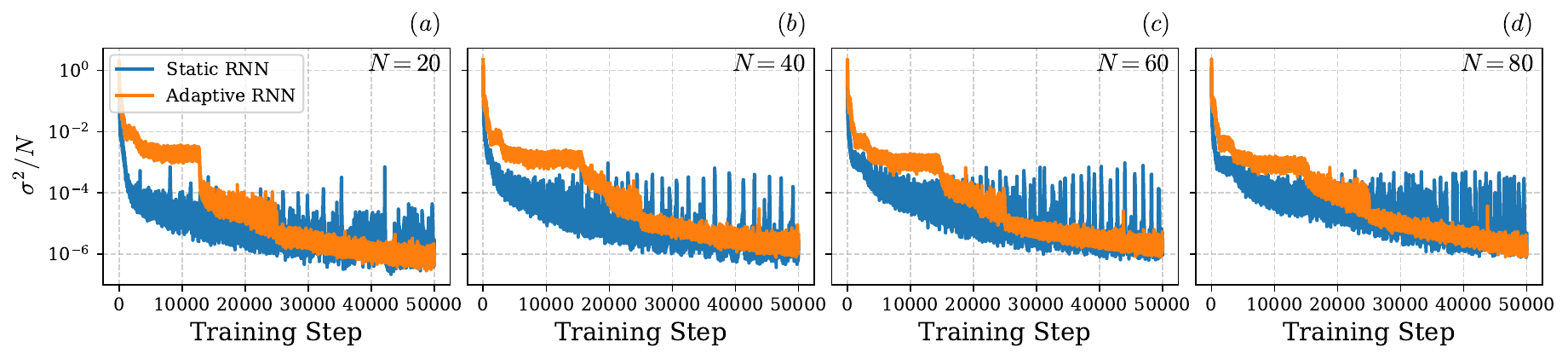}
    \caption{Variance per spin throughout training for the 1D TFIM benchmark across all system sizes not shown in manuscript. (a) $N=20$, (b) $N=40$, (c) $N=60$, (d) $N=80$.}
    \label{fig:AllSizes}
\end{figure*}

In this Appendix, we highlight the strong fluctuations in the Static RNN compared to the Adaptive RNN in terms of the energy variance per spin during training, as demonstrated in Fig.~\ref{fig:AllSizes}. This result highlights the enhanced stability of training provided by the Adaptive training scheme. By decreasing the learning rate, the fluctuations are reduced. However, higher energy variances are obtained.

\bibliography{Biblio}
\end{document}